\documentstyle[preprint,revtex,eqsecnum]{aps}
\begin{document}
\draft
\begin{title}
Hawking radiation: a particle physics perspective.
\end{title}
\author{Matt Visser\cite{email}}
\begin{instit}
Physics Department, Washington University, St. Louis,
Missouri 63130-4899
\end{instit}

\receipt{20 April 1992}

\begin{abstract}
It has recently become fashionable to regard black holes as elementary
particles. By taking this suggestion seriously it is possible to
cobble together an elementary particle physics based estimate for
the decay rate $(\hbox{black hole})_i \to (\hbox{black hole})_f +
(\hbox{massless quantum})$. This estimate of the spontaneous emission
rate contains two free parameters which may be fixed by demanding
that the high energy end of the spectrum of emitted quanta match
a blackbody spectrum at the Hawking temperature. The calculation,
though technically trivial, has important conceptual implications:
(1) The existence of Hawking radiation from black holes is ultimately
dependent only on the fact that massless quanta (and all other
forms of matter) couple to gravity. (2) The thermal nature of the
Hawking spectrum depends only on the fact that the number of internal
states of a large mass black hole is enormous.  (3) Remarkably,
the resulting formula for the decay rate  gives meaningful answers
even when extrapolated to low mass black holes.  The analysis
strongly supports the scenario of complete evaporation as the
endpoint of the Hawking radiation process (no naked singularity,
no stable massive remnant).
\end{abstract}

\pacs{04.20.-q, 04.20.Cv, 04.60.+n; Wash U HEP/92-73; hepth@xxx/9204062}
\narrowtext
\newpage
\section{INTRODUCTION}

It has recently become clear that the statistical description of
the Hawking radiation process breaks down at low entropy \cite{PSSTW}.
Buoyed by this and other results a number of authors have investigated
the utility of considering low mass black holes as elementary
particles \cite{Holzhey-Wilczek,Giddings-Strominger,Giddings}.

In the present paper, I propose taking the particle physics aspects
of black hole physics seriously. Due to the absence of any
fundamental theory of quantum gravity I will have to resort to
physical reasoning to motivate an estimate for the decay rate
\begin{equation}
(\hbox{black hole})_i \to
(\hbox{black hole})_f + (\hbox{massless quantum})
\end{equation}
For definiteness, one may take the massless quantum to be a photon,
though neutrinos or linearized gravitons would do as well.  A major
goal of this paper will be to motivate Hawking radiation from a
particle physics perspective, minimizing geometrodynamic aspects
of the problem.

The estimate for $\Gamma_0(b_i \to b_f + \gamma)$ that I shall cobble
together has several striking features:\\
(1) In the limit of a large black hole, where statistical techniques
will be shown to be appropriate, the model predicts emission of a
Maxwell--Boltzmann spectrum with two free parameters.
These two parameters may be fixed by matching with Hawking's
semiclassical calculation which is valid in this parameter regime
\cite{Hawking1,Hawking2}.\\
(2) If the black hole interacts with a bath of photons, stimulated
emission occurs. The spontaneous emission rate (appropriate for a
black hole emitting radiation into a vacuum) is modified: $\Gamma
= (1+\langle n \rangle)\Gamma_0$. If the external photon bath is
in equilibrium with itself at some temperature $T_\gamma$ this
enhancement factor gives $\Gamma(T_H,T_\gamma) = \{
e^{\hbar\omega/T_\gamma} / (e^{\hbar\omega/T_\gamma} - 1) \}
\Gamma_0(T_H)$. If furthermore the photon bath is also in equilibrium
with the black hole itself, ($T_\gamma = T_H$), then the emission
and absorption spectra are equal to each other and are exactly
Planckian.\\
(3) Once the two free parameters have been fixed in this manner,
the resulting model continues to give sensible physical results
even when extrapolated to small mass black holes --- strongly
suggesting that the final stage of the Hawking radiation process
is a cascade of gammas (and other massless quanta) leading to
complete evaporation with no naked singularity and no stable massive
remnant.

\underbar{Units:} $c=1$; $k=1$; $G=\ell_P/m_P = \ell_P^2/\hbar =
\hbar/m_P^2$.

\section{THE MODEL}
\subsection{Particle physics}

Consider the two body decay $b_i \to b_f + \gamma$. The initial
and final black holes are parameterized by mass $M$, spin $J$, and
possess internal states whose number is denoted by $N$. The interaction
vertex is just a photon coupling gravitationally to matter.
Furthermore the decay is prohibited neither by phase space
considerations nor by any conservation law. Therefore by general
particle physics principles this decay {\it must} take place. The
only question is at what rate. Without further ado, consider the
estimate
\begin{equation}
\hbar \Gamma(b_i \to b_f + \gamma) = \xi
      \left( {\hbar \omega \over m_P} \right)^2
      \left( {A_H\over \ell_P^2} \right)
      \left( \hbar \omega \right)
      \left(2\;\; {2J_f+1\over2J_i+1}\right)
      { N_f \over N_i }.
\end{equation}
The various terms in this estimate have the immediate physical
interpretation:\\
\underbar{(1) coupling constant:} The dimensionless coupling constant
describing matter coupling to gravity is $g = E/m_P$. The decay rate
depends on the square of the coupling constant.\\
\underbar{(2) surface enhancement:} The emitted photon can couple
to the black hole at any point on its horizon. Since the coupling
is gravitational one can plausibly argue that different pieces of
the horizon, of order a Planck length squared in area, will couple
independently and incoherently to the emitted radiation. This
implies a decay rate proportional to the area of the event horizon.\\
\underbar{(3) phase space:} The volume of two particle phase space
is proportional to $\hbar \omega$.\\
\underbar{(4) spin factors:} These arise from a sum over squares
of Clebsch--Gordon coefficients.\\
\underbar{(5) statistical factors:} One should average over initial
states and sum over final states. The number of internal states of
a black hole of mass $M$ has been denoted $N(M)$.\\
An overall dimensionless numerical factor $\xi$ is left undetermined
by this argument.

To reiterate the basic ingredient of this estimate: If a decay is
(a) kinematically allowed, (b) does not violate conservation laws,
and (c) the initial and final states are coupled, then that decay
{\it will} and in fact {\it must} take place.

\subsection{Kinematics}

Consider the two body decay $b_i \to b_f + \gamma$. Work in the
rest frame of the initial state. Elementary kinematics yields the
energies of the final state particles
\begin{equation}
\hbar\omega = {M_i^2 - M_f^2 \over 2 M_i}; \qquad
E_f = {M_i^2 + M_f^2 \over 2 M_i} = \sqrt{M_f^2 +(\hbar\omega)^2}.
\end{equation}

\subsection{Maxwell--Boltzmann spectrum}

With malice aforethought, one defines $N=e^S$.  Suppose now that
for $M_{(i/f)} >> m_P$, one can show $N_{(i/f)} >> 1$. Then $S_{(i/f)}
>>1$, and one may adopt a statistical approach, treating $M_{(i/f)}$,
$N_{(i/f)}$, $S_{(i/f)}$ and $\hbar\omega$ as effectively continuous
variables.

Combining this statistical approximation with two body kinematics gives
for the initial-final state factor
\begin{eqnarray}
{N_f\over N_i} \equiv \exp\left(S_f - S_i\right)
               &\approx& \exp\left(
	                   {\partial S \over \partial M} [M_f-M_i]
	                   \right) \nonumber \\
               &\approx& \exp\left( -
	                   {\partial S \over \partial M} \hbar \omega
	                   \right)
               \equiv \exp(-\hbar\omega/T).
\end{eqnarray}
Where, with further malice aforethought, one defines $T^{-1} \equiv
(\partial S/\partial M)|_{M_i}$. At this stage this is merely a
definition but it is instructive to see that the gross features of
a thermal spectrum are already emerging without recourse to curved
space quantum field theory, Euclidean continuations of the metric,
or similar arcane.

Up to this point, the argument of this paper has provided the estimate
\begin{equation}
\Gamma_0(b_i \to b_f + \gamma) = 2\xi
      \left( {\hbar \omega \over m_P} \right)^2
      \left( {A_H\over \ell_P^2} \right)
      \omega \;
      \left( {2J_f+1\over2J_i+1} \right)
      \exp(-\hbar\omega/T).
\end{equation}
Perhaps more interestingly, the decay rate into the frequency {\it
interval} $[\omega,\omega+d\omega]$ can be written as
\begin{equation}
d\Gamma_0(b_i \to b_f + \gamma) =  2\xi
      \left( {\hbar \omega \over m_P} \right)^2
      \left( {A_H\over \ell_P^2} \right)
      \left( {2J_f+1\over2J_i+1} \right)
      \exp(-\hbar\omega/T) \; d\omega.
\end{equation}
Therefore, the power radiated into this frequency range is
\begin{equation}
dP(\omega) =  \hbar\omega \; d\Gamma_0
           = 2\xi A_H
            \left( \hbar \omega \right)^3
            \left( {2J_f+1\over2J_i+1} \right)
            \exp(-\hbar\omega/T) \; d\omega,
\end{equation}
which is the Maxwell--Boltzmann spectrum promised in the introduction.
The two parameters ($\xi$, $T$) are now fixed by appealing to the
known behaviour of the Hawking radiation in the high frequency
limit. Taking $J_i=J_f=0$, comparison with Hawking's original
calculation \cite{Hawking1,Hawking2} or any of a number of subsequent
calculations \cite{incantations,textbook} shows that $\xi = 1/(3\pi)$
and $T=T_H$.

\subsection{Stimulated emission}

Suppose now that instead of radiating into a vacuum, the black hole
is radiating into a bath of photons. If a certain mode is already
inhabited by $n$ photons it is well known that the rate $n\to n+1$
is enhanced by stimulated emission: $\Gamma_{n\to n+1} = (1+n)
\Gamma_{0\to1}$. After averaging over $n$ the total emission rate
is
\begin{equation}
\Gamma
= \sum_n p_n \Gamma_{n\to n+1}
= (1+ \langle n \rangle) \Gamma_{0\to1}.
\end{equation}
Now if the photon bath is in equilibrium with itself at some
temperature $T_\gamma$  Bose statistics implies that $\langle n \rangle =
[\exp(\hbar\omega/T_\gamma) - 1]^{-1}$. The total emission rate for
the black hole is then
\begin{equation}
\Gamma(T_H,T_\gamma) =
{ e^{\hbar\omega/T_\gamma} \over e^{\hbar\omega/T_\gamma} - 1 }
\Gamma_0(T_H)
\end{equation}
If furthermore the photon bath is also in equilibrium with the
black hole itself, ($T_\gamma = T_H$), then the emission and
absorption spectra are equal to each other and are both exactly
Planckian.
\subsection{The Bekenstein entropy estimate}

{}From an analogy between the classical law of area growth for event
horizons and and the second law of thermodynamics, Bekenstein
\cite{Bekenstein} argued that the physical entropy of a (large)
black hole is
\begin{equation}
S = {1\over4\eta} {A_H\over\ell_P^2}.
\end{equation}
Here $\eta$ is an undetermined numerical constant. Equivalently,
the Bekenstein estimate for the number of internal states of a
(large) black hole is
\begin{equation}
N= \exp\left( {1\over4\eta} {A_H\over \ell_P^2} \right).
\end{equation}
By appealing to either curved space quantum field theoretic
calculations \cite{Hawking1,Hawking2} or to manipulations involving Euclidean
analytic continuations of the manifold \cite{incantations},
Bekenstein's estimate may be verified, and the constant $\eta$
found to be $\eta = 1$.  For any reasonably large black hole, this
immediately implies $N>>1$, thereby justifying the statistical
approximation adopted above.

\subsection{Comments}

This calculation, though technically trivial, teaches one several
very important things. (1) The existence of Hawking radiation from
black holes is ultimately dependent only on the fact that photons
(and all other forms of matter) couple to gravity. (2) The thermal
nature of the Hawking spectrum depends only on the fact that the
number of internal states of a large mass black hole is enormous.
(3) The formal definitions $S=\ln N$ and $T^{-1} \equiv (\partial
S/\partial M)$ really do correspond to the physical entropy and
temperature of the black hole. (4) These conclusions are insensitive
to the precise nature of the black hole: Schwarzschild,
Reissner--Nordstrom, Kerr--Newman, or dilatonic. (5) Curved space
quantum field theory calculations are required only when one wishes
to  explicitly calculate  the entropy (or, equivalently, the number
of internal states) as a function geometrodynamic parameters.

\section{THE ISSUE OF THE FINAL STATE}
\subsection{Generalities}

One of the most intriguing aspects of this microscopic modelling
of the Hawking radiation process is that the resulting formalism
continues to give meaningful and quite physical answers when
extrapolated to small black holes (where the statistical and
semiclassical arguments usually employed fail utterly).

All the evidence assembled up to this point indicates that one
should take the notion of the number of internal states of a black
hole seriously --- as though $N$ were a real physical parameter.
Proceed by {\it really} taking this notion of the number of internal
states seriously. One notes that the number of internal states is
by definition an integer. One infers a discrete spectrum for the
allowed entropy.
\begin{equation}
S(N) =  \ln N.
\end{equation}
For a general black hole, the entropy will be related to the mass,
and to other parameters such as charge, spin, and the presence or
absence of a dilaton field. The quantization of entropy implies a
quantization of mass $M(N)$, with the precise form of $M(N)$
depending on the type of black hole in question.  The spontaneous
decay rate is then
\begin{equation}
\Gamma_0(b_i \to b_f + \gamma) =
      {2 \over 3\pi}
      \left( {M(N_i)^2 - M(N_f)^2 \over 2 \hbar M(N_i)} \right)^3
      A_H(N_i) \; {2J_f+1\over2J_i+1} \;{ N_f \over N_i }.
\end{equation}
For a small black hole this implies the onset of a cascade process
with characteristic time scale of order the Planck time. When does
one expect this formula to break down? If the emitted photons have
energy greater than or of order the Planck mass, then their Compton
wavelength is less than their ``Schwarzschild radius'', and one
would expect the emitted photons to ``disappear down their own
event horizons'' rendering the present estimate unreliable.
Conversely, if one can show that the energy gaps are small
compared to the Planck scale, and if the average energy of emitted
photons is small compared to the Planck scale, then there is no
obstruction to asserting the validity of this formula down to the
lowest mass scales.

One may normalise $N$ so that $N=1$ corresponds to the lowest mass
black hole of the given type.  Then the final decay from the $N=1$
state should be handled by means different from the above.  For
instance, for the lowest mass neutral black hole, consider the
decay into two gammas $b_1 \to \gamma + \gamma$. One may easily
estimate
\begin{equation}
\Gamma(b_1 \to \gamma + \gamma)
= \zeta \left({M_1\over2m_P}\right)^4 \left({A\over\ell_P^2}\right)^2
{M_1\over2\hbar}.
\end{equation}
The only novelty in this estimate is that the amplitude is second
order in the coupling constant $g= (M_1/2)/m_P$, so that the decay
rate is fourth order. Additionally each photon can couple independently
somewhere on the horizon so there are two area enhancement factors.
Naturally the dimensionless constant $\zeta$ is not calculable by
present techniques. From a variation of the discussion in the
previous paragraph, this particular estimate is of course trustworthy
if and only if $(M_1/2) << m_P$. Again, I wish to emphasise: (a)
photons couple to gravity, (b) kinematics allows the decay, and
(c) no conservation laws are violated, therefore the decay {\it
must} take place. These ideas generalize: For instance, for the
lowest mass singly charged black hole $b_1^+$ the same sort of
analysis can be applied {\it mutatis mutandis} to the decay $b_1^+
\to e^+ \bar{\nu_e}$.

This analysis rather strongly suggests that the endpoint of the
Hawking evaporation process is a cascade of gammas (and other light
particles), emitted as the black hole descends to the $N=1$ state,
followed by complete evaporation into two light particles (no naked
singularities, no stable massive remnants).

\subsection{Schwarzschild black holes}

Consider now an ordinary unadorned Schwarzschild black hole. The
mass spectrum alluded to previously is easily seen to be
\begin{equation}
M^2(N) = m_P^2 \; {\ln(1 + N) \over 4\pi}.
\end{equation}
For large $N$, the mass gap between neighbouring states is of order
$\delta(M^2) = m_P^2/N$, so that (as expected) the mass of large
mass black holes is effectively a continuous parameter.  The
spontaneous decay rate is
\begin{equation}
\Gamma_0(b_i \to b_f + \gamma)
      = {4 \over 3} (4\pi)^{-5/2} \omega_P
      \left[ \ln\left({1+N_i\over1+N_f}\right) \right]^3
      \left[\ln(1+N_i)\right]^{-1/2}
      { N_f \over N_i }.
\end{equation}
It may easily be verified that all of the emitted gammas have energy
much less than the Planck mass. The highest energy gamma from this
cascade arises in the $(N=2) \to (N=1)$ decay for which $\hbar\omega
= m_P \bigg[\sqrt{(\ln3)/4\pi}-\sqrt{(\ln2)/4\pi}\bigg] = (0.0608...)
m_P$.

The putative final decay into two gammas $b_1 \to \gamma + \gamma$,
produces photons of energy $E=(1/2)M_1 = (1/2)\sqrt{(\ln2)/4\pi}
\; m_P = (0.1174...) m_P$, which is indeed safely smaller than the
Planck scale.

The lesson to be learned is this: from a particle physics perspective
the evaporation of a Schwarzschild black hole can be plausibly tracked
all the way to the final state without ever having to directly
confront Planck scale physics.

\subsection{Kerr-Newman black holes}

Once the mass of a Schwarzschild black hole drops below $M\approx
m_P^2/m_e$, the Hawking temperature rises above the mass of an
electron $T_H > m_e$.  Once this happens the Hawking radiation will
include electrons and positrons as well as photons (and gravitons
and neutrinos). Since electrons and positrons carry off both spin
and charge, it is clear that any truly realistic description of
black hole evaporation cannot be modelled solely using Schwarzschild
geometry, but must at some level include the effects of charge and spin.

Consider a charged, spinning black hole described by the Kerr--Newman
geometry. Adopt units wherein electric charge is measured in mass
units, {\it ie} $q^2/(4\pi\epsilon_0) \equiv GQ^2$, so that the
charge on an electron is $Q_e \equiv \sqrt{\alpha} m_P$. I shall
for the time being content myself with writing down the spectrum.
Starting from the standard result for the entropy \cite{PSSTW}
\begin{equation}
S = {2\pi M^2\over m_P^2} \left\{ 1 - {Q^2\over2M^2}
                   +\sqrt{1-{Q^2\over M^2}-{L^2\over G^2 M^4} } \right\},
\end{equation}
straightforward manipulations yield
\begin{equation}
M^2 = m_P^2 {\pi\over S}
      \left\{
         \left({S\over2\pi} +{Q^2\over2m_P^2} \right)^2 + {L^2\over\hbar^2}
      \right\}.
\end{equation}
The mass spectrum is then simply
\begin{equation}
M^2(N,Z,J) = m_P^2 {\pi\over\ln(1+N)}
      \left\{
         \left({\ln(1+N)\over2\pi} +{Z^2\alpha\over2} \right)^2 + J(J+1)
      \right\}.
\end{equation}
Note that $J=Z=0$ reproduces the Schwarzschild case, that $N=Z=J=0$
is the vacuum state, and that $N=0$ with $Z \neq 0 \neq J$ is
forbidden.  Verifying that the mass gaps are small is an exercise
in tedious integer algebra.

After spitting out a cascade of gammas, charged leptons, neutrinos,
quarks, {\it etc}, the black hole will minimize its mass by eventually
settling to a $N=1$, $J=0$ state (possibly with $Q\neq0$). The most
obvious manner in which the black hole might then lose its electric
charge, emission of an electron, is energetically forbidden since
it requires the black hole to increase its spin. Thus subsequent
evolution depends on whether or not the particle physics spectrum
contains any charged scalars. If elementary charged scalar particles
exist then there is no barrier to the black hole losing its electric
charge by spitting out these charged scalars, thereby descending
to the $N=1$, $J=Z=0$ state considered previously. On the other
hand, if charged scalars do not exist, the lowest order process
leading to charge loss is the semi--gravitational semi--weak three
body decay $b_i^Z \to b_f^{Z-1} + e^+ + \bar{\nu_e}$ which proceeds
via a virtual $W^+$. This process will eventually reduce the black
hole to the $N=1$, $Z=1$, $J=0$ state, where the two body decay
$b_1^+ \to e^+ \bar{\nu_e}$ can take over. Again, one sees the same
basic physics in operation: Gravity couples to everything, thus,
when kinematics and superselection rules permit the decay, that
decay {\it will} take place.

The only hope for a stable massive remnant is if the decay is
forbidden by a conservation law. For instance, if the black hole
is a magnetic monopole, and if no ``ordinary'' particles carry
magnetic charge, then the lowest mass black hole corresponding to
a given magnetic charge must be stable.

\subsection{Dilatonic black holes}

Consider a spinning charged axionic dilatonic black hole.  Such a
black hole is a solution of the combined gravitational, electromagnetic,
axion and dilaton equations of motion.  The spin zero case has been
discussed by Gibbons and Maeda \cite{Gibbons-Maeda}, and more
recently by Garfinkle, Horowitz, and Strominger \cite{GHS}. Further
variations on the spin zero theme may be found in references
\cite{IY1989,IY1992,STW,CKO,Lee-Weinberg}. Generalizations to
nonzero spin were first explored by Horne and Horowitz
\cite{Horne-Horowitz}, and the complete solution worked out by Sen
\cite{Sen}. The entropy of such a black hole is
\begin{equation}
S= {2\pi M^2 \over m_P^2}
   \left\{1 - (Q^2/2M^2) +\sqrt{[1-(Q^2/2M^2)]^2 - (L^2/G^2 M^4)}\right\}.
\end{equation}
As a function of entropy and charge,
\begin{equation}
M^2 = {Q^2\over2} +
      {m_P^2} \left\{{S\over4\pi} +{\pi L^2\over S\hbar^2 } \right\}.
\end{equation}
One easily infers the mass spectrum
\begin{equation}
M(N,Z,J)^2 = m_P^2
           \left\{ {Z^2\alpha\over2} +
	           {\ln(1+N)\over4\pi} +
		   {\pi J(J+1)\over\ln(1+N)}
	   \right\}.
\end{equation}
This class of black holes is in fact just as easy to analyse as
the Kerr--Newman case. Note that $J=Z=0$ reproduces the Schwarzschild
case, that $N=Z=J=0$ is the vacuum state, and that $N=0$ with $Z
\neq 0 \neq J$ is forbidden.  As for the Kerr--Newman geometry,
verifying that the mass gaps are desirably small is an exercise in
tedious integer algebra. The discussion of the decay cascade may
be carried over {\it mutatis mutandis}.

\section{DISCUSSION}

In the absence of a reliable theory of fully quantized gravity the
problem of dealing with the final stages of Hawking evaporation
via an {\it ab initio} calculation appears to be a completely
hopeless task.  The arguments presented in this note seek to sidestep
the whole issue by developing a phenomenological particle physics
based model for the Hawking evaporation process. The model developed
herein adequately accounts for the Hawking radiation process from
heavy black holes --- albeit with two free parameters that have to
be fixed by comparison with Hawking's semiclassical calculation.

Having obtained the model, the model yields finite meaningful
results even when extrapolated to small black holes. The model
rather strongly supports the notion that the final state of the
Hawking radiation process is complete evaporation. It appears that,
from a particle physics perspective, the evaporation of a black
hole can be plausibly tracked all the way to the final state without
ever having to directly confront Planck scale physics.  Because
the model treats Hawking radiation as a cascade of quite ordinary
particle physics decays, the model automatically preserves quantum
coherence. Any initially pure quantum state will evolve to some
pure (but complicated) final state.

Insofar as the ideas presented in this paper make any sense it now
becomes absolutely critical to face head on the question: ``What
exactly {\it are} the internal states of a black hole?''.

\acknowledgements

I wish to thank Carl Bender, Adrian Burd, Mike Ogilvie, and Don
Petcher for their interest and comments.  This research was supported
by the U.S. Department of Energy.



\begin{references}
\bibitem[*]{email}Electronic mail: visser@kiwi.wustl.edu
\bibitem{PSSTW}
J. Preskill, P. Schwartz, A. Shapere, S. Trivedi, and F. Wilczek,\\
Mod. Phys. Lett. {\bf A6}, 2353 (1991).
\bibitem{Holzhey-Wilczek}
C. F. E. Holzhey and F. Wilczek, \\
Black holes as elementary particles,\\
IASSNS-HEP-91/71, hepth@xxx/9202014, December 1991.
\bibitem{Giddings-Strominger}
S. B. Giddings and A. Strominger,\\
Dynamics of extremal black holes,\\
UCSB-TH-92-01, hepth@xxx/9202004, February 1992.
\bibitem{Giddings}
S. B. Giddings,\\
Black holes and massive remnants,\\
UCSB-TH-92-09, hepth@xxx/9203059, March 1992.
\bibitem{Hawking1}
S. W. Hawking,
Nature {\bf 248}, 30 (1974).
\bibitem{Hawking2}
S. W. Hawking,
Commun. Math. Phys. {\bf 43}, 199 (1974).
\bibitem{incantations}
G. W. Gibbons and S. W. Hawking,
Phys. Rev. {\bf D15}, 2752 (1977).
\bibitem{textbook}
N. D. Birrell and P. C. W. Davies,\\
{\it Quantum fields in curved space},\\
(Cambridge Press, Cambridge, 1982).
\bibitem{Bekenstein}
J. Bekenstein,
Phys. Rev. {\bf D7}, 2333 (1973).
\bibitem{Gibbons-Maeda}
G. W. Gibbons and K. Maeda,
Nucl. Phys. {\bf B298}, 741 (1988).
\bibitem{GHS}
D. Garfinkle, G. T. Horowitz, and A. Strominger,
Phys. Rev. {\bf D43}, 3140 (1991).
\bibitem{IY1989}
I. Ichinose and H. Yamazaki,
Mod. Phys .Lett. {\bf A4}, 1509 (1989).
\bibitem{IY1992}
H. Yamazaki and I. Ichinose,
Class. Quantum Gravit. {\bf 9}, 257 (1992).
\bibitem{STW}
A. Shapere, S. Trivedi, and F. Wilczek,
Mod. Phys. Lett. {\bf A6}, 2677 (1991).
\bibitem{CKO}
B. A. Campbell, N. Kaloper, K. A. Olive
Phys. Lett. {\bf B263}, 364 (1991).
\bibitem{Lee-Weinberg}
K. Lee and E. J. Weinberg,
Phys. Rev. {\bf D44}, 3159 (1991).
\bibitem{Horne-Horowitz}
J. T. Horne and G. T. Horowitz,\\
Rotating Dilaton Black Holes,\\
UCSB-TH-92-11, hepth@xxx/9203083, March 1992.
\bibitem{Sen}
Ashoke Sen,\\
Rotating charged black hole solution in heterotic string theory,\\
TIFR/TH/92-20, hepth@xxx/9204046, April 1992.
\end{references}
\end{document}